%% file: conference_101719.tex
\pdfoutput=1

\documentclass[conference]{IEEEtran}
\IEEEoverridecommandlockouts
\usepackage{tikz}
\usepackage{pgfplots}
\pgfplotsset{width=10cm,compat=1.9}

\usetikzlibrary{arrows.meta,
                chains,
                positioning}

\usepackage{cite}
\usepackage{amsmath,amssymb,amsfonts}
\usepackage{algorithmic}
\usepackage{graphicx}
\usepackage{textcomp}
\usepackage{verbatim}
\usepackage{xspace}
\usepackage{xcolor}
\usepackage{fontawesome}
\usepackage{lookup}
\usepackage[utf8]{inputenc}
\usepackage{color, colortbl}
\usepackage{lmodern}
\usepackage{tcolorbox}
\usepackage{balance}
\usepackage{url} 
\usepackage{hyperref}

\def\BibTeX{{\rm B\kern-.05em{\sc i\kern-.025em b}\kern-.08em
    T\kern-.1667em\lower.7ex\hbox{E}\kern-.125emX}}
\begin{document}

\title{Please Turn Your Cameras On: Remote Onboarding of Software Developers during a Pandemic}

\author{\IEEEauthorblockN{Paige Rodeghero\IEEEauthorrefmark{1}\IEEEauthorrefmark{2}, Thomas Zimmermann\IEEEauthorrefmark{2},
Brian Houck\IEEEauthorrefmark{2}, and
Denae Ford\IEEEauthorrefmark{2}}
\IEEEauthorblockA{\IEEEauthorrefmark{1}School of Computing, Clemson University\\
Clemson, South Carolina 29631 \\
Email: prodegh@clemson.edu}
\IEEEauthorblockA{\IEEEauthorrefmark{2} Microsoft Corp. \\
Redmond, WA 98052\\
Email: \{tzimmer, brian.houck, denae\}@microsoft.com}}

\maketitle
\input{aliases.tex}

\begin{abstract}
The COVID-19 pandemic has impacted the way that software development teams onboard new hires.  Previously, most software developers worked in physical offices and new hires onboarded to their teams in the physical office, following a standard onboarding process.  However, when companies transitioned employees to work from home due to the pandemic, there was little to no time to develop new onboarding procedures.  In this paper, we present a survey of 267 new hires at Microsoft that onboarded to software development teams during the pandemic.  We explored their remote onboarding process, including the challenges that the new hires encountered and their social connectedness with their teams. We found that most developers onboarded remotely and never had an opportunity to meet their teammates in person.  This leads to one of the biggest challenges faced by these new hires, building a strong social connection with their team. We use these results to provide recommendations for onboarding remote hires.
\end{abstract}


\input{participants_lookup}

\input{introduction}
\input{background}

\input{onboardingremotely}
\input{onboardingremotelyresults}
\input{discussion}
\input{conclusion}

\section*{Acknowledgments}

We thank and acknowledge Tony Hernandez for his discussion with us about his experience of onboarding new hires remotely to software development teams.  We also thank Andrew Begel and Simina Pasat for discussing their previous experience researching software developers' onboarding experience at Microsoft and GitHub. We also thank Caroline Davis, Susan Hastings Tiscornia, Kanwal Safdar, Tanya Platt, Angie Anderson, and Kelly Sieben for the privacy and ethics review of these studies.

\balance
\bibliographystyle{IEEEtran}
\bibliography{IEEEabrv,biblio}

\end{document}

%% file: aliases.tex
\definecolor{darkgreen}{rgb}{0.05,0.5,0.05}

\newcommand{\pr}[1]{{\color{blue}\bfseries Paige: #1}}
\newcommand{\df}[1]{{\color{darkgreen}\bfseries Denae: #1}}
\newcommand{\tz}[1]{{\color{orange}\bfseries Tom: #1}}
\newcommand{\someone}[1]{{\color{red}\bfseries TODO: #1}}

\newcommand{\covid}{COVID-19\xspace}

\renewenvironment{quote}{%
   \list{}{%
     \leftmargin\parindent%
     \rightmargin0cm
   }
   \item\relax
}
{\endlist}

\newcommand{\myquote}[2]{\begin{quote}{\small\faComment}~\emph{#1} (\lookupGet{#2})\end{quote}}
\newcommand{\myquotex}[1]{\begin{quote}{\small\faComment}~\emph{#1}\end{quote}}

\newcommand{\inlinequote}[2]{\emph{``#1''} (\lookupGet{#2})\xspace}
\newcommand{\inlinequotex}[1]{\emph{``#1''}}

%% file: participants_lookup.tex
\lookupPut{R_ZvL4CfDnZbNe11L}{S1}
\lookupPut{R_RIyfRtxsQqlzgKl}{S2}
\lookupPut{R_xnhz3Y9Rm71sTdv}{S3}
\lookupPut{R_3L4HiNTwJdsfyfk}{S4}
\lookupPut{R_RlyzjZLrdA5TZDP}{S5}
\lookupPut{R_2Xh6Y0gkMgA8u6h}{S6}
\lookupPut{R_3KJhLzV24Wz7EUd}{S7}
\lookupPut{R_3nvI3Gz5kVoU2V7}{S8}
\lookupPut{R_2XclQiWYdMAUrqy}{S9}
\lookupPut{R_yvVfriEcMguDx9n}{S10}
\lookupPut{R_Dk7rZF3zrsAM7IJ}{S11}
\lookupPut{R_PAnNnAbnxDtSvRv}{S12}
\lookupPut{R_1Iic1t14cogAty2}{S13}
\lookupPut{R_2Yh9zWfolGlyCgk}{S14}
\lookupPut{R_TckFQW7YDDMthSx}{S15}
\lookupPut{R_xaY4j46vqquENlT}{S16}
\lookupPut{R_1OVRZKduIXYTmAk}{S17}
\lookupPut{R_2Qsq9M7NevuGdfB}{S18}
\lookupPut{R_3JDjSHqp2qpxIRe}{S19}
\lookupPut{R_bK4tmePEasbLNTz}{S20}
\lookupPut{R_b3FgHeffNM5Botz}{S21}
\lookupPut{R_svAED6NdlZh9oK5}{S22}
\lookupPut{R_D7ufCN1HX0F1JF7}{S23}
\lookupPut{R_2xGkbJN3airBqH4}{S24}
\lookupPut{R_1Cvc4eEsmbtux6C}{S25}
\lookupPut{R_1mJoe5MBQv3VIf6}{S26}
\lookupPut{R_BM6iIo2kpSadTbP}{S27}
\lookupPut{R_1rlHEHaypqfUKOH}{S28}
\lookupPut{R_2QR204ozM9AbC2d}{S29}
\lookupPut{R_2CPycrl4GrSa8VG}{S30}
\lookupPut{R_tYxaXCOEn1MQEo1}{S31}
\lookupPut{R_2CTcJJCcIB9ymB9}{S32}
\lookupPut{R_z2VHUayRdgKziUh}{S33}
\lookupPut{R_3MsecvZHuHwDEuJ}{S34}
\lookupPut{R_1MMA3ERo37OZ3EW}{S35}
\lookupPut{R_1GKdC46wNrtJUBh}{S36}
\lookupPut{R_1OkUmk7B3gldbXB}{S37}
\lookupPut{R_pi9OSfwUceQcAA9}{S38}
\lookupPut{R_2rMylZA828gG2XF}{S39}
\lookupPut{R_3PMHPOWt4nptNHk}{S40}
\lookupPut{R_2Ys017xpORgMMju}{S41}
\lookupPut{R_3FKAuS7m29ZkeMG}{S42}
\lookupPut{R_1Owbz4hpYS9OSIF}{S43}
\lookupPut{R_3kpKwxXcFxw5PxH}{S44}
\lookupPut{R_3nMiXSu3jGTAXlD}{S45}
\lookupPut{R_sz19gMR1vFnqV7r}{S46}
\lookupPut{R_RyG4iJUl5EUTGWR}{S47}
\lookupPut{R_1cZLrLXMCfAa4hT}{S48}
\lookupPut{R_3lqkDbio6ifsxj3}{S49}
\lookupPut{R_1dy1rJl5xmWVYdn}{S50}
\lookupPut{R_6QGZodIMnyrdX7b}{S51}
\lookupPut{R_2D2KxGY9oytNPK1}{S52}
\lookupPut{R_27TCS7bTATZrXjx}{S53}
\lookupPut{R_2s6jLg0fiaj331C}{S54}
\lookupPut{R_sFiWSYB5iKUJRlL}{S55}
\lookupPut{R_XU5c1vYdYlKOBjj}{S56}
\lookupPut{R_POJxBzoTuWYyQdb}{S57}
\lookupPut{R_1mDwOCufRiULHAW}{S58}
\lookupPut{R_3JEEY4JelHc6to7}{S59}
\lookupPut{R_3HCgNAsZFJO73yW}{S60}
\lookupPut{R_3fVpmhDXAvIvlOJ}{S61}
\lookupPut{R_1oiE8V5gULk1AwA}{S62}
\lookupPut{R_2CPaDFEyJjfH59P}{S63}
\lookupPut{R_XXJmffXePbnrAg9}{S64}
\lookupPut{R_2tJaSpbuctNmGKv}{S65}
\lookupPut{R_2pVNbgbjyP1VQQN}{S66}
\lookupPut{R_8v8pU1jRzGtyhLb}{S67}
\lookupPut{R_2xKlDG5eWlO2rTK}{S68}
\lookupPut{R_2ZENK2v6z0tN3mp}{S69}
\lookupPut{R_qI6Qby8TpE3YxfX}{S70}
\lookupPut{R_1rDpZSfUUKnYV6E}{S71}
\lookupPut{R_1fj8o5SCKU8TL4P}{S72}
\lookupPut{R_30vfkApn2z66dIf}{S73}
\lookupPut{R_cBXI2WdG0641Q0F}{S74}
\lookupPut{R_1QikhjcptyetDzc}{S75}
\lookupPut{R_08JXjftJIArLVdL}{S76}
\lookupPut{R_3CK2YO2tuIcZrbs}{S77}
\lookupPut{R_2sRQYxb3Qa0VXAa}{S78}
\lookupPut{R_3nkGw1cGPRGuNe5}{S79}
\lookupPut{R_3MGKDulq84fBe6X}{S80}
\lookupPut{R_1PbYxG7wZAxKnWB}{S81}
\lookupPut{R_30f8NYGPbFxxA2k}{S82}
\lookupPut{R_3QW7T3EPAJfVJnY}{S83}
\lookupPut{R_12yDdFqV67y9W1O}{S84}
\lookupPut{R_27B5p4VHFCS0gDc}{S85}
\lookupPut{R_3mjBTKU12R79wGR}{S86}
\lookupPut{R_1q4LpSwtF5BsNJe}{S87}
\lookupPut{R_3s6AUbUqDN1I7yI}{S88}
\lookupPut{R_27xUlAYHFfi5dDJ}{S89}
\lookupPut{R_yQIQLZUCTRfUSu5}{S90}
\lookupPut{R_2Bmp55vvz6eaTrs}{S91}
\lookupPut{R_1ozr76kBdiRbmw2}{S92}
\lookupPut{R_7NXFj1vWxZ2RNU5}{S93}
\lookupPut{R_bPKLVW36f9BfP0J}{S94}
\lookupPut{R_3kMQpvNbieSanPw}{S95}
\lookupPut{R_1q24LLulSjd0z4q}{S96}
\lookupPut{R_2bZEgZ9oTj3Ubmx}{S97}
\lookupPut{R_3PpFMpzmEYwRumE}{S98}
\lookupPut{R_9zy5952CBzvqGfD}{S99}
\lookupPut{R_XhsydS3WKlIWJot}{S100}
\lookupPut{R_1K2cEcqoSL5bLQG}{S101}
\lookupPut{R_1gRy9NUP6mCdoz4}{S102}
\lookupPut{R_2wgZApHbRBqdBKU}{S103}
\lookupPut{R_1OIvOp5Z2VzXlWr}{S104}
\lookupPut{R_3p8Vu3fsKShGfFG}{S105}
\lookupPut{R_3dRZgj5ki3Um2Nn}{S106}
\lookupPut{R_BzVsLMZ2NhkBxYJ}{S107}
\lookupPut{R_3dYex4kHx3p9166}{S108}
\lookupPut{R_0CHhlxw0w343Tbz}{S109}
\lookupPut{R_27drmcbYy7gIscj}{S110}
\lookupPut{R_2AZps9YDDMyQHM2}{S111}
\lookupPut{R_D7Tp1iCAk5RRRHH}{S112}
\lookupPut{R_1ItS5o6xaW7yNMT}{S113}
\lookupPut{R_sh7zGJpCHMY4DD3}{S114}
\lookupPut{R_3HAixtqJ7jWu20j}{S115}
\lookupPut{R_2RaISz3v0J1cmVm}{S116}
\lookupPut{R_DOvOdeYsQG7kJpv}{S117}
\lookupPut{R_QfC5APqYiRo2Eet}{S118}
\lookupPut{R_2BgMMx46LBksoV5}{S119}
\lookupPut{R_qwkUIwtv96uADp7}{S120}
\lookupPut{R_1ieZJN3xiNvWlBe}{S121}
\lookupPut{R_Zge1zNQpzgHgXcJ}{S122}
\lookupPut{R_1K70RjjJ9HRn3f5}{S123}
\lookupPut{R_3PTfsGhmosUhhG5}{S124}
\lookupPut{R_32QnQHohArVE1uR}{S125}
\lookupPut{R_1Km6tiN42Za7YXN}{S126}
\lookupPut{R_9XYzllSLlVHEnoR}{S127}
\lookupPut{R_Oy6dWLH9NGKOPsd}{S128}
\lookupPut{R_3Dp8buerylfl9lr}{S129}
\lookupPut{R_sXUhHGP9abHa11L}{S130}
\lookupPut{R_27vTsOqjuYQQUIj}{S131}
\lookupPut{R_tVApz0mrgNXw4X7}{S132}
\lookupPut{R_21cWQywKPJKMJVA}{S133}
\lookupPut{R_z6a4Is8CU2WESLn}{S134}
\lookupPut{R_2tFJYvLGNkbVi6C}{S135}
\lookupPut{R_2EacNhiN7cfGbJE}{S136}
\lookupPut{R_R35FTxFFPaEmUQF}{S137}
\lookupPut{R_33mxF0yjfpte6yN}{S138}
\lookupPut{R_POIbi523uylxq01}{S139}
\lookupPut{R_2fBcard96t4fWGA}{S140}
\lookupPut{R_1HcZRKkmupfF5hT}{S141}
\lookupPut{R_1DG89AFooQfUwdj}{S142}
\lookupPut{R_24cMysYB9Q3ePzP}{S143}
\lookupPut{R_w0Ose84YpNT4iMp}{S144}
\lookupPut{R_1mwrqopUTXmf14R}{S145}
\lookupPut{R_10wJreZcRWmBkqN}{S146}
\lookupPut{R_2rVojVd6OtsS025}{S147}
\lookupPut{R_4Orwxti9NDlj5MB}{S148}
\lookupPut{R_3gSKZu6U2VXoy9w}{S149}
\lookupPut{R_6Y8HMk3bZKNXMD7}{S150}
\lookupPut{R_1GP5Cy2JBU3DpzB}{S151}
\lookupPut{R_3hbEgEbNbfdCQ64}{S152}
\lookupPut{R_3xeliDG6eGfVv2h}{S153}
\lookupPut{R_z0evoEkw3kPkL3H}{S154}
\lookupPut{R_2BajeB0LcCpbY3h}{S155}
\lookupPut{R_3PitdY4QilQnGjp}{S156}
\lookupPut{R_1EWfNovbVAtd3Bh}{S157}
\lookupPut{R_CetYKF1Tzer891v}{S158}
\lookupPut{R_1gtJS00DOuD8KcD}{S159}
\lookupPut{R_1pSyMuQXB70gME4}{S160}
\lookupPut{R_2eaRhLPQt4OkoEN}{S161}
\lookupPut{R_3JaRy26Y9WsbWUt}{S162}
\lookupPut{R_2bHR4nBCBZaZMWr}{S163}
\lookupPut{R_1mWA0Co4fnI4YgN}{S164}
\lookupPut{R_bCpFfsIx7li9dEl}{S165}
\lookupPut{R_2rTIVw9WRrl8B6T}{S166}
\lookupPut{R_3MF4NWTYcTkTLV5}{S167}
\lookupPut{R_XtvRZsKX06j0cGB}{S168}
\lookupPut{R_3R2ARsXhNRquZeU}{S169}
\lookupPut{R_1HqxX2JESSMlXCo}{S170}
\lookupPut{R_2UYqeDXlAdZPSkc}{S171}
\lookupPut{R_eIHZCzFYp4pFpVT}{S172}
\lookupPut{R_OcCxBb4GQXOM64x}{S173}
\lookupPut{R_3POcQnUIqe3kcxk}{S174}
\lookupPut{R_3svTXi8Z4UwDGJe}{S175}
\lookupPut{R_1N47gHZsLJvE1pQ}{S176}
\lookupPut{R_2du7G0bAiY6GPWt}{S177}
\lookupPut{R_2zuzU7Li3MXXC5Q}{S178}
\lookupPut{R_2aDclr6Q1mNWjwx}{S179}
\lookupPut{R_DuXdl5NhReIJCal}{S180}
\lookupPut{R_2Xp5HROH0J4ISGN}{S181}
\lookupPut{R_1pJQFkRKOpR4aCh}{S182}
\lookupPut{R_1HkjA0wjqNyNci1}{S183}
\lookupPut{R_2tbVuO30VCL2gaz}{S184}
\lookupPut{R_56DLsG9QTFuGJ6p}{S185}
\lookupPut{R_1NF8KKerA5npe17}{S186}
\lookupPut{R_31KuozsRf8RNKQ8}{S187}
\lookupPut{R_2D2MYQHPYX3wcQv}{S188}
\lookupPut{R_3rE5N0PMhfl74g9}{S189}
\lookupPut{R_cUa9WFHS3aufFwB}{S190}
\lookupPut{R_3QQvqf0EDN95Jzy}{S191}
\lookupPut{R_30v2exinmpWklbM}{S192}
\lookupPut{R_3lWcXt7enJblO6l}{S193}
\lookupPut{R_27UoNAG2tZ2ZC80}{S194}
\lookupPut{R_NVDVBj8ZkhirVg5}{S195}
\lookupPut{R_2fkcMse6Tl2h2mY}{S196}
\lookupPut{R_1kMpWFMLPHnSaua}{S197}
\lookupPut{R_2uzKW6zoRum4nyp}{S198}
\lookupPut{R_O3jIEcH0OCuOgFP}{S199}
\lookupPut{R_cGZtSRcJh2930Hf}{S200}
\lookupPut{R_3Ev7TAxw3mlr5Fw}{S201}
\lookupPut{R_3ma2TIW3vvnLStX}{S202}
\lookupPut{R_2ahJYJDuwPzRu7U}{S203}
\lookupPut{R_1OOlDO5V1jRB3uS}{S204}
\lookupPut{R_DUgN43tMNpmB7sR}{S205}
\lookupPut{R_1NrZguU7OBE9dfA}{S206}
\lookupPut{R_1jOeQgiSyNmTVrs}{S207}
\lookupPut{R_22YsccTwDsqQwI9}{S208}
\lookupPut{R_1ewnTjVyuhsLHK1}{S209}
\lookupPut{R_1IsfmdkQTBJf4XP}{S210}
\lookupPut{R_AnRqFryla9B7F7z}{S211}
\lookupPut{R_3F4JhXGy4gYajSc}{S212}
\lookupPut{R_9GjYkPKoWcPmXx7}{S213}
\lookupPut{R_20TcHv3hix2SZDK}{S214}
\lookupPut{R_1EYbmKS4EqDBC36}{S215}
\lookupPut{R_3oReY789I0FIHe3}{S216}
\lookupPut{R_2YXngAQhpUsmKAo}{S217}
\lookupPut{R_2OZ8N5CqEWf02eQ}{S218}
\lookupPut{R_2ZEyo0x5abq3Dxa}{S219}
\lookupPut{R_2xDl1oWcevMLp0e}{S220}
\lookupPut{R_ZCNfsDnRgRR2tsl}{S221}
\lookupPut{R_3315gBQX0xOx3bG}{S222}
\lookupPut{R_3e8F1NrFuCtcA7j}{S223}
\lookupPut{R_3qUlA2LSPytxUzp}{S224}
\lookupPut{R_0qtzYI3d9yyxqqB}{S225}
\lookupPut{R_2ZCsWnFKfF077st}{S226}
\lookupPut{R_214lM30SBCcuo0b}{S227}
\lookupPut{R_3HOfQ3o7fbU3HnJ}{S228}
\lookupPut{R_20Ya5yCAQouWAKF}{S229}
\lookupPut{R_3EmreSclvTz1gcQ}{S230}
\lookupPut{R_dh5wKmDrljU5LJn}{S231}
\lookupPut{R_1QxZUQKQQpyDygM}{S232}
\lookupPut{R_1BR7dRKfqfWog9j}{S233}
\lookupPut{R_1hRwZ27TujGpJf6}{S234}
\lookupPut{R_1QDDRJnEiFWejg7}{S235}
\lookupPut{R_A1hgHgLEak7ywY9}{S236}
\lookupPut{R_3sb0mtX9cIEgCRZ}{S237}
\lookupPut{R_DpdryxEKpuCmEUh}{S238}
\lookupPut{R_089ObNr7iu5zF5v}{S239}
\lookupPut{R_1jZz1kZlIzEAyOQ}{S240}
\lookupPut{R_VKZmIbuj7C9n5x7}{S241}
\lookupPut{R_895nT1VBB1fIzjr}{S242}
\lookupPut{R_db4Es9yeCGh9InD}{S243}
\lookupPut{R_3nf72e8wAGmhSlB}{S244}
\lookupPut{R_3ezsOUwT98Kn8bq}{S245}
\lookupPut{R_ZJDIxWwWt7vyCiZ}{S246}
\lookupPut{R_27g36xLMhWUoKNj}{S247}
\lookupPut{R_bNkzxKRLNJFjF2V}{S248}
\lookupPut{R_sczqax3Z4tEA88x}{S249}
\lookupPut{R_UZ9Sl3wqYZhL1eh}{S250}
\lookupPut{R_3mkvpBQrhBUK6QB}{S251}
\lookupPut{R_2SlhOJt8TIorAO6}{S252}
\lookupPut{R_247Oa5KOVAWrRwj}{S253}
\lookupPut{R_1eXY30NlBYDa0VO}{S254}
\lookupPut{R_1o7muYhz7vUsYQi}{S255}
\lookupPut{R_4ZwjbSXKeJabnJ7}{S256}
\lookupPut{R_3j7kRgDxgbUVfNQ}{S257}
\lookupPut{R_2ty4bttEmku7AmT}{S258}
\lookupPut{R_2tmSG2ixP1jlwH0}{S259}
\lookupPut{R_2Ev1CbVOVwJvQAQ}{S260}
\lookupPut{R_Y60Fo8OLqTDRNUR}{S261}
\lookupPut{R_3PcXJn7Sd0VIohw}{S262}
\lookupPut{R_1P7z6mfqBArhg3I}{S263}
\lookupPut{R_1LMQCoJNqSd9ekJ}{S264}
\lookupPut{R_2ePA8AbxqPgFiE8}{S265}

%% file: introduction.tex
\section{Introduction}
``Onboarding'' is the period of time when programmers become familiar with a new project, its source code, and the team.  The onboarding process is a ``necessary evil.'' This process is undesirable due to its extreme expense and error proneness~\cite{labuschagne2015onboarding, bonar1983uncovering}. In response to the expensive onboarding process, the software engineering research community has responded by experimenting with methodologies, such as pair programming~\cite{williams2004initial}, to overcome the barriers for onboarding.  Most of the research efforts have been focused on the human effort~\cite{ fagerholm2014role, steinmacher2015social, labuschagne2015onboarding}.  For example, mentoring during the onboarding process has been a heavily researched area.  Companies, such as Google, have tried to improve the onboarding process using mentorship programs~\cite{johnson2010learning}. Sim et al. found programmers had a shallow understanding of projects they were working on even after four months~\cite{sim1998ramp}.  Even worse, Zhou and Mockus found that programmers need a minimum of three years to become fluent in a project~\cite{zhou2010developer}.  It is important to note that the average turnover rate at a large company such as Amazon or Google is less than two years~\cite{peterson_2017}. This means that many programmers may never be fully onboarded.

Unfortunately, the onboarding techniques described in the literature are largely not applicable to programmers onboarding remotely. ``Remote'' is commonly used to describe a programmer that works away from their physical office~\cite{hill1998influences}.  They may work at home, in a shared public space, or any location of their choice. The idea of remote work is that a programmer can work in any chosen location and effectively contribute to a development team.  Remote onboarding presents unique challenges in comparison to the local onboarding because new hires work off-site.

The traditional workplace changed overnight when the \covid pandemic spread~\cite{chowell2020covid}.  Employees transitioned to work from home (WFH), unless considered an essential employee by the government (e.g., a respiratory therapist). %
Many developers were hired by Microsoft before the lock-down but had not officially started their new job. Once starting their job, most of these new hires went through a remote onboarding process without meeting their team in person or even stepping into the office. %

At the time of submitting this research paper, we do not have an end date for the \covid pandemic.  Furthermore, and unfortunately, we do not know how long the pandemic will last.  However, we do know that the future of work has changed.  For example, Twitter and Square found that remote work was sufficient and announced that employees could continue to WFH indefinitely~\cite{brownlee_2020}. This means that remote onboarding will grow of importance in the future as other companies move to remote work.
Without multiple studies exploring the onboarding challenges that new remote hires encounter, we do not know how best to tailor the remote onboarding process. Most of the field has become remote for the conceivable future. 

\begin{tcolorbox}[width=0.48\textwidth]
\underline{\textbf{The Problem:}}\\
In this paper, we address the following gap in the current onboarding literature: there are no research studies on how software engineers onboard to a team during a pandemic. This gap currently presents a problem for software development companies hiring and onboarding new employees.  
\end{tcolorbox}

In this paper, we focus on improving the onboarding process during the COVID-19 pandemic. We surveyed 267 new hires that remotely onboarded to their teams due to WFH.  We explored their onboarding process, their connection with their team, and the challenges they encountered while onboarding.  We found that the majority of the new hires felt welcomed and supported by their team.  We also found that many new hires felt moderately connected with their team, even though they may not have had interaction with their team every day.  Finally, we found that the challenges new hires experienced while onboarding included technology issues, finding useful documentation, communication and collaboration difficulties, and difficulties building a strong connection with their team.  To conclude this paper, we provide extensive recommendations for managers and teams onboarding new hires remotely and during a crisis.

%% file: background.tex
\section{Background and Related Work}\label{back}
In this section, we cover the background and related work on onboarding developers and remote work during the \covid pandemic.

\subsection{Onboarding Software Developers}
Onboarding is a key area for software companies' success because of the knowledge-intensive nature of work involved with complex software projects~\cite{britto2018onboarding}. The reasons for onboarding new developers can be from replacing retired developers, replacing developers who left or will leave the organization, scaling up the number of developers, or freeing up other developers to work on other tasks ~\cite{britto2018onboarding}. While there is significant literature in general onboarding ~\cite{gruman2006organizational, klein2015specific, lynch2010reducing, johnson2005get, moyson2018organizational}, there is a lack of studies involving best practices and strategies for the onboarding of remote hires.  Additionally, there is no research on onboarding published during the pandemic, a time where the work has shifted overnight into a new direction. Before the pandemic, researchers had conducted case studies to explore the onboarding of new hires at Google~\cite{johnson2010learning}.%
The work studied Google's new employees' onboarding process used benchmarks to achieve desired goals and reduce isolation and increase new employees' collegiality~\cite{johnson2010learning}. 

In 2008, Hemphill and Begel studied remote onboarding at a major software development company~\cite{begel2008pair}.  They conducted empirical case studies on five software development teams within the company that had recently hired and onboarded their first remote team member. The authors conducted direct observations and interviews with the managers, new hires, and mentors.  They found that the new remote hires desired a stronger social connection to their team. Additionally, the new remote hires struggled to understand the work process and find the needed information when they had questions.  Specifically, they found a lack of documentation that the new hires needed to understand their team's work process their team follows.   

In another holistic study by Britto ~\textit{et al.}, three separate case studies were conducted to investigate the long-term and short-term onboarding outcomes. These outcomes were grouped into six different categories. They include recruiting, orientation, training, coaching and supper, support tool and processes, and feedback. The results from the case studies include recommendations and implications for best practices while onboarding developers. These results include the following:  Explaining the expectations for new hires during the recruitment process, have a written onboarding plan that lays out objects, timelines, roles, and responsibilities. Transparency must be provided into the organization's projects and the key roles other developers have across multiple locations of the organization. Allowing developers to travel to met others if the project is distributed across remote locations.  Providing in-depth coaching and support for legacy projects and tailoring each onboarding program to suit the developers' needs. It is important to note that are limitations to the different toolsets each organization has available \cite{britto2018onboarding}. Finally, another study of the remote onboarding process of remote teams concluded that remote onboarding processes were almost non-existing because of the lack of opportunity for collocated interactions \cite{hemphill65not}. Our work further explores the remote onboarding process and we provide recommendations for improving a multitude of areas of that process, including the ways that teams interact and bond.

\subsection{Remote Work and Remote Work during the Pandemic}
 Even before the pandemic, remote work had a host of advantages such as work autonomy, reduced interruptions, and flexible schedules~\cite{ford2019remotework}. The advantages that remote work provides have inspired many technologies, such as Stack Overflow and GitLab, to support their employees being remote-first~\cite{mazzina2017remotefirst} or even all-remote~\cite{gitlab2020remotereport}. In fact, this trend had become so popular that many companies started providing reports and writing guides to help other organizations to do the same~\cite{gitlab2020playbook}. These materials cite recommendations to help make remote work a priority for organizations, such as providing new employees with a dedicated remote buddy whom they can go to with any question~\cite{pardue2017remotework}. Another recommendation is to have social events to connect everyone (across remote and in-person settings) intentionally. These pre-pandemic responses on how to support remote work highlight the importance of support, which is definitely important given the pandemic stage of work we are currently in.

Although developers are working from home, it is important to note that remote work in a pandemic is not the same as traditional remote work (before \covid)~\cite{hanselman_2020}.
There are also a host of new challenges that the pandemic brings.  With COVID-19, many people are socially isolated (e.g., living alone)~\cite{usher2020life}.  There are no opportunities to travel to the office and limited opportunities to form a social connection with the team.  There is also a new challenge for many of those with children.  Many people have to juggle homeschooling in addition to completing their technical work~\cite{cluver2020parenting}. Not to mention, traditional remote work did not consider many experiencing trauma while still working~\cite{horesh2020traumatic}. Since the pandemic, studies of software developers have found that there is truly a dichotomy of experiences impacting productivity~\cite{bao2020does,  ralph2020pandemic}. For example, while some people find it easier to find more uninterrupted focus time since no colleagues are stopping by their office, others are finding it difficult to track down and communicate with colleagues~\cite{ford2020tale}. 
But what about developers who are new to the team, let alone the company? Our work investigates the challenges new developers face when onboarding remotely to a new company and team.

%% file: onboardingremotely.tex
\section{Remote Onboarding Survey Design}
In this section, we describe our research questions and the methodology of our onboarding survey.

\subsection{Research Questions}\label{rqs}
Our goal is to better understand how remote developers onboard during a pandemic and their challenges.  Towards this goal, we highlight four areas of onboarding and remote work that previous studies have explored before the pandemic (See Section II).  We study these areas by posing the following Research Questions (RQs):   

\begin{description}
\item[$RQ_{1}$] What are the challenges that developers encountered when onboarding during the pandemic?  
\item[$RQ_{2}$] What are teams doing to onboard new hires?
\item[$RQ_{3}$] How do team members interact with new hires during the pandemic?
\item[$RQ_{4}$] Do new hires feel socially connected to their team? 

\end{description}

The rationale behind  $RQ_1$ is that no software development company has onboarded employees during a pandemic.  Software development companies did not have an onboarding process or plan for when a pandemic would happen. Similar to many software development companies, Microsoft had hired software developers before WFH and those new hires expected to onboard in the physical office.  Unfortunately, due to the lockdown, most of those new hires remotely onboarded from their home.  We ask this research question in order to understand better the challenges that these developers faced while onboarding remotely.  Next, we ask $RQ_2$ to explore the efforts that teams made towards remotely onboarding new employees during their own transition to WFH. We ask $RQ_3$  to understand the types of interactions that occurred between the new hires and their teammates.  Finally, we ask  $RQ_4$ to understand the new hires and their team's social connection. Social connection is essential for team productivity~\cite{reagans2001networks,yilmaz2016effective} and job satisfaction~\cite{ayazlar2014effect, lopes2005informal}.  Also, we ask this question because studies in organizational psychology literature report challenges in building trust and connection between teammates in distributed teams~\cite{dery2016seeing}. %

\input{figures/timeline-teamproductivity}

\subsection{Methodology}
To understand the onboarding experience of new hires during the \covid pandemic, we designed and deployed an online survey within Microsoft (see Figure~\ref{fig:study-timeline}). Microsoft employees in the location of the company headquarters (King County) were instructed to work from home on March 4, 2020. 
The design of the survey was based on the related work on onboarding and software development during \covid (see Section \ref{back}).   We used piloting to test earlier versions of the survey.
The final online survey consisted of 50 questions that we deployed using Qualtrics and distributed in the first weeks of July. %
Our questions explored the new hires' onboarding experience, their interaction with their team during the onboarding process, their perceived productivity, and the overall degree of connection they felt with their team. Most questions were optional in the survey.
For this paper, we analyzed a subset of the questions asked.

\smallskip
\subsubsection*{Participants} We sent recruitment emails to 1000 new software engineering hires who started at Microsoft between January and June 2020.  We received 267 responses to the survey (26.7\% response rate). 
The respondents were professional developers, all new hires, and full-time employees at Microsoft.  
Of the participants, 57 identified as women, 162 participants as men, and one participant identified as non-binary/gender diverse. (47 participants did not respond to the gender question).
The participants were a mix of hire types: 140 were industry hires, 14 were industry hires with previous experience working at Microsoft, 38 of the participants were college hires, and 26 had recently graduated from college and previously interned or consulted at Microsoft. %
We only surveyed employees in the United States due to the lock downs and reopenings occurring at different times throughout the world.  All of the participants surveyed were working from home. Once participants completed the survey, they could enter into a sweepstakes to win one of four \$100 USD Amazon gift certificates.

\smallskip
\subsubsection*{Data Analysis}
For the open-ended survey questions, we followed an open-coding approach to analyze the responses~\cite{mayring2004qualitative,spencer2004card}. Three of the authors coded a random sample of the data for each question. Once the data was coded, the authors discussed the resulting codes and themes and through negotiated agreement~\cite{garrison2006revisiting} compiled a code book for each question. This approach allowed us to find commonalities in the responses and form themes, which we will discuss in the later sections of this paper.

\subsection{Limitations / Threats to Validity}

Any empirical study has limitations. First, our participants onboarded in different weeks, during different months, to various teams throughout North America. This means that they interacted with different managers, and the onboarding process that employees went through is most likely unique from one another. This also means that the participants filled out the survey during different stages of their onboarding process. A survey could control for this aspect in normal times, for example, by sending out the survey invitation three months after each new hire's start date. We considered such a survey design, however, it would not have been able to control for the different stages of the pandemic as the lock downs rules have often been evolving on a weekly basis. Instead, we decided to contact all new hires at the same time (first weeks of July) during the pandemic.

Second, we studied only employees at a single company. Employees at other companies may have a different onboarding experience, especially if the company is different in size or had a remote work culture before the pandemic. However, by studying new hires in a single company, as opposed to new hires in many different companies, we could control for aspects such as the company's pandemic response, which would be different across companies. We are confident that most of our recommendations are useful for most software companies.

Third, it is possible that the survey questions could be subjective because an employee wrote the survey.  To help mitigate this, the first author was brought in as a visiting researcher and was not a full-time or permanent employee.

Fourth, the new hires may not have noticed the efforts made by their teams.  This could bias our results. We do note that most team efforts were unique to their teams. However, our observations were similar across teams. 

Finally, due to the stress of the pandemic, our results may be impacted by outside stress.  Once the pandemic is over, this outside stress will hopefully disappear. While pandemic stress may have influenced the challenges reported in this paper, we know that the future of work has been changed, and the current experiences will impact how remote onboarding will take place in the future.

%% file: figures/timeline-teamproductivity.tex
\begin{figure}[!htb]
\centering

\resizebox{\linewidth}{!}{%
\begin{tikzpicture}[
  node distance = 1mm and 3mm,
  start chain = A going below,
  dot/.style = {circle, draw=white, very thick, fill=gray, minimum size=3mm},
  box/.style = {rectangle, text width=65mm, inner xsep=4mm, inner ysep=1mm,
                on chain},
  ]
\begin{scope}[every node/.append style={box}]
\node { \emph{Employees in King County instructed to work from home}} ; 
\node { \emph{Most US states under stay at home orders (94\% of the US population)} } ;
\node { \textbf{Survey} sent to a total of 1,000 full-time software engineers in the United States. \textbf{267 responses (response rate: 26.7\%)}} ;
\end{scope}
\draw[very thick, gray, {Circle[length=3pt]}-{Triangle[length=4pt)]},
      shorten <=-3mm, shorten >=-3mm]   
      (A-1.north west) -- (A-3.south west);
\foreach \i [ count=\j] in {\emph{March 4, 2020},\emph{April/early May},July 1--July 17}
    \node[dot,label=left:\i] at (A-\j.west) {};
\end{tikzpicture}
}
\caption{Timeline of survey}
\label{fig:study-timeline}
\end{figure}
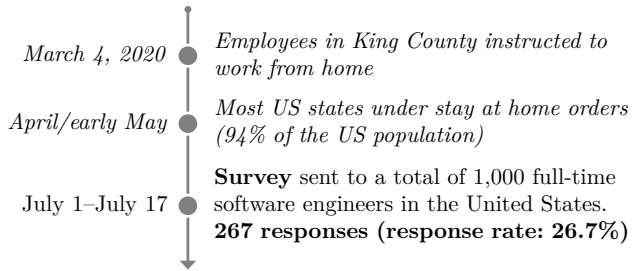

%% file: onboardingremotelyresults.tex
\section{Remote Onboarding Survey Results}
We present our answers to each research question and our data, rationale, and interpretation of the answers.

\newcommand{\questionNumber}[1]{\textbf{#1}\xspace}
\renewcommand{\questionNumber}[1]{}

\subsection{Pandemic Onboarding Challenges ($RQ_1$)}\label{challenges}%
We first asked participants \textbf{``What challenges have you faced while onboarding to your team and project?''} \questionNumber{(Q26)} We followed up the open-ended question with a question to ask the participants if the challenges reported were specific to COVID-19 \questionNumber{(Q27)} and 61.9\% of the participants who responded, selected ``Definitely yes'' or `Probably yes.''%

\newcommand{\myparagraph}[1]{\medskip\noindent\textbf{#1}\xspace}

We found that participants had various responses to the first question about what challenges they faced during the onboarding process.  We found six common onboarding challenges that emerge from the data:   

\myparagraph{Communication and Collaboration.} Communication and collaboration challenges were a common theme in the data. By communication, we mean connecting with teammates for scheduled and unscheduled meetings or chat messages.  By collaboration, we mean brainstorming and working together.   
     
     \myquote{During meetings, most teammates keep the camera off. This makes collaboration and remote experience difficult.}{R_3HOfQ3o7fbU3HnJ}
     \myquote{%
There could've been more communication in the first 2-3 weeks for me to be more productive.}{R_3j7kRgDxgbUVfNQ}
     \myquote{Not enough communication due to WFH}{R_1mDwOCufRiULHAW}

\myparagraph{Asking for Help.} We found that participants struggled to ask for help while working remotely. We found that the reasons for this varied from isolation between teammates due to the nature of remote work, scheduling difficulties, the lack of running into peers in the hallway, etc. 
        \myquote{starting remotely has certainly been a disadvantage in getting to know people enough to be comfortable with randomly interrupt them multiple times a day for help.
}{R_3MsecvZHuHwDEuJ}
         \myquote{Because everyone is working from home I don't run into people randomly or walk by their office, which has led me to ask for less help than I usually would have because I don't like asking small, half-thought out questions virtually.}{R_3dRZgj5ki3Um2Nn}
            \myquote{I haven't been able to get enough help or specific feedback on the tasks I'm working on. I can ask questions and talk but quick emails and chats don't give me enough clarity. I lack confidence that my work is actually what it's supposed to be. I've had to scrap some things and do them over differently because there was accidental miscommunications and that's made me move forward more slowly now. }{R_2Qsq9M7NevuGdfB}

\myparagraph{Building Team Connections.}
We found that respondents often mentioned that they struggled to build connections within their team.
\myquote{Onboarding remotely is definitely more challenging.  The inability to truly interact with coworkers, the inability to really have traditional team building, and the disconnection from everyone contributes to feeling like my work is kind of isolated from everyone else.  It also makes it harder to get a broader, high level understanding of the team and organization.}{R_2Bmp55vvz6eaTrs}
Respondents also discussed barriers to building a connection to their team.  Respondents noted that when their team did not use their videos in calls, it created a challenge in understanding their new team's dynamics, and it was hard to bond with them.  %
\myquote{This being a remote joining, it is difficult to connect to the team and not having daily stand ups or a weekly team meeting where the videos are on makes it difficult to know the team dynamics. Also, there is no teams channel where all the communication happens where a new hire can at least watch and learn (something that is similar to listening to peers talk if we were in office) instead of the team having one to one conversations on private messaging}{R_08JXjftJIArLVdL}

\myquote{...People generally are not comfortable using the video during meetings (plus video takes up more bandwidth), so it rarely is used. As a result it has been challenging to get to know people.}{R_1GKdC46wNrtJUBh}In general, we found that new hires commonly noted that onboarding and working remotely were very difficult for building a connection with their team.
\myquote{Onboarding remotely is definitely more challenging.  The inability to truly interact with coworkers, the inability to really have traditional team building, and the disconnection from everyone contributes to feeling like my work is kind of isolated from everyone else.  It also makes it harder to get a broader, high level understanding of the team and organization. Switching to remote right when I started working on feature development in a new language and environment. Support for learning was all of a sudden much harder.}{R_2Bmp55vvz6eaTrs}
\myquote{It's hard to gauge if I'm really meshing with the team working remotely. It's also a little bit harder to get help with questions because I can't just walk to someone's desk.}{R_2XclQiWYdMAUrqy}

\myparagraph{Finding Documentation.} New hires reported difficulty finding needed resources (e.g., wikis). 
   
   \myquote{There was not enough documentation on the projects I've been working on. I had to ask a lot of questions to my coworkers to be able to work efficiently. Most of these questions could have been written in some documentation.}{R_2xKlDG5eWlO2rTK}
    Additionally, some new hires reported documentation missing, which caused delays in starting work. 
   
    \myquote{Setting up the development environment had taken me close to 2 weeks and I feel there should have been a stackoverflow sort of internal tool to help us out on common issues. These common issues have not been documented anywhere and so more often than not a team member had to help me out. I feel teams should have proper onboarding guide/documentation detailing the steps to setup the environment and also steps to verify if everything is working.}{R_1Iic1t14cogAty2}
   Others reported excessive information spread out in multiple locations.
   
  \myquote{A lot of information spread out in too many places. Not easy to find certain information}{R_3kpKwxXcFxw5PxH}
 
  \myquote{...there is both an information glut and an indexing dearth.}{R_1PbYxG7wZAxKnWB}New hires found that some documentation was outdated, a common complaint among the majority of software developers in industry~\cite{garousi2013evaluating,lethbridge2003software}.  

  \myquote{Some of the onboarding documentation was outdated.}{R_sXUhHGP9abHa11L}Finally, we found that participants reported that their challenge onboarding was searching for the internal documentation itself.
\myquote{It is difficult to search for internal wikis and documentation. This slowed the velocity of onboarding a lot.}{R_2rVojVd6OtsS025}

\myparagraph{Technical Issues.} New hires found that they encountered technical issues.  The technical issues ranged from challenges with setting up development environments to issues with remote restarts.
       
       \myquote{At first issues with Bitlocker and restarting my computer, which I had to do a number of times in the beginning. I didn't realize you could suspend your protection and restart it.}{R_3kMQpvNbieSanPw}

\myparagraph{Hardware and Permissions.} Some new hires reported difficulty in quickly acquiring the needed hardware for their job because they had to have the equipment shipped to them. We found that this challenge prevented them from starting work quickly.  
    
\myquote{Getting all of the software/hardware I need while being remote. Sometimes I need to wait for hardware to be delivered. Other times I need to get software that I can't easily access from home. I often don't know what software I need until someone can get back to me. }{R_3JDjSHqp2qpxIRe}

\subsection{Team Onboarding Efforts ($RQ_2$)}%
Teams have had to re-imagine their efforts when onboarding new hires due to WFH. We asked participants \textbf{``What has the team done to make you feel connected to the team?''} \questionNumber{(Q17)}
We found that some new hires had frequent 1:1 meetings with their manager. 
\myquote{My manager did 1:1 every single day for first two weeks. There was introduction email send out broad audience. Also had introduction meeting.
There is teams chat group for regular conversation. Overall it was great experience.}{R_UZ9Sl3wqYZhL1eh}
We also found that multiple teams held a remote meeting for new hires to meet their new team on their first day.
\myquote{The team has been extremely welcoming. My very first day we met over teams and each person introduced themselves and welcomed me personally. Several people have reached out to ask how my onboarding is going and offered to answer any questions I have. Several members of my team invited me to join them in a hackathon. Generally everyone has been incredibly helpful and patient when teaching me.}{R_3QQvqf0EDN95Jzy}
Some participants noted their team's efforts to ensure they were not blocked on a task and assisted in networking.
\myquote{They constantly check if I need anything. They are assigning me the work such that I will be able to connect with different people in the team.}{R_30v2exinmpWklbM}
We also found that teams were holding social virtual events to bond and get to know one another.  These events varied from a casual reoccurring meeting to playing online games with one another.
\myquote{The team has set up a reoccurring meeting to just relax and chat with each other. It is a nice event for me to meet and start to get to know my teammates better.}{R_32QnQHohArVE1uR}
Finally, we found that teams assigned onboarding buddies to assist in the onboarding process. 
\myquote{My manager assigned a buddy within my team. I connect with him more often. Also, about 4 of us in the team meet 3 days a week to share our status. I work closely with them and they are very responsive in Teams.}{R_3JaRy26Y9WsbWUt}
\myquote{I had an onboarding buddy which was very helpful. I still go to him for onboarding type questions (they still come up), especially if there is a tool or resource I run into that I haven't used yet.}{R_1GKdC46wNrtJUBh}
We found that 151 (67.7\%) respondents were assigned a mentor and 72 were not assigned a mentor. \questionNumber{(Q30)}

\subsection{Team Interaction ($RQ_3$)}%
We found that the interactions varied between new hires and their teams. For example, 129 (58.1\%) new hires reported having 1:1 meetings daily with someone from their team, while 93 new hires did not have daily 1:1s with other team members. \questionNumber{(Q39)}%
We asked participants \textbf{``What does your typical daily informal and formal communication with your team look like?''} \questionNumber{(Q38)}We found similar responses from new hires to this question. 
\myquote{It's all formal communication via Teams.  Every once in a while a small subset of us newer engineers will banter with each other.}{R_OcCxBb4GQXOM64x}
\myquote{Formally there are scheduled meetings - mostly with the whole team, but I do have a scheduled 1:1 with my manager. There are informal chats and some email exchanges between me and individuals to help me.}{R_2Qsq9M7NevuGdfB}

We also asked \textbf{``What does that communication look like? What content is typically discussed?''} \questionNumber{(Q40)}We found that most communication occurred using Microsoft Teams and the majority of the communication is focused on work related topics. We followed this question up by asking \textbf{``On average, how long does the communication last?''} \questionNumber{(Q41)}We found that 72.9\% of the reported communication lasted less than 30 minutes and 43.4\% of communication was less than 15 minutes.

\subsection{Social Connection to Team ($RQ_4$)} %
When asked \textbf{``How socially connected do you feel to your team?''} we found that 83.3\% of the new hires reported a connection with their team. \questionNumber{(Q34)}To learn more about the strategies of new hires to stay socially connected, we additionally asked \textbf{``How do you stay socially connected with your team?''}
Many respondents expressed excitement over team bonding activities. %

\myquote{We have a weekly `game night' where we meet and play online games together. It's probably the favorite part of my week!}{R_3nvI3Gz5kVoU2V7}
We found a few instances where the new hires mentioned their efforts to main social connection during technical meetings with their team and peers.

\myquote{I try to reach out to random team members. I'm soon going to ask my manager if people can start turning on their cameras so I can see facial expressions.}{R_1QikhjcptyetDzc}

\myquote{When opportunities arise (such as 1:1 technical discussions) I try to deviate from the work related item under discussion.}{R_z2VHUayRdgKziUh}
A few new hires reported struggles with staying connected with their team due to the WFH orders.

\myquote{It's been difficult to stay socially connected with my team due to WFH situation. There has been some effort recently to increase social interactions by setting up time to vent about any issues we are facing with teammates.}{R_2D2KxGY9oytNPK1}
Finally, a few survey respondents had not yet made actively efforts to stay connected with their teammates.  
\myquote{I met with my manager in person twice. Beside that I don't stay socially connected with others.}{R_27UoNAG2tZ2ZC80}
\myquote{Unfortunately I guess I haven't taken active steps to do so since I feel like I'm spending all of my time trying to ramp up and learn.}{R_1ozr76kBdiRbmw2}

\subsection{Summary of Onboarding Survey Results}
We derive three key interpretations from our qualitative and quantitative results.  First, software developers faced many challenges while onboarding remotely to a software development team during the \covid pandemic.  Some of these challenges, such as a lack of documentation, are not unique to the remote work setting.   However, other challenges, such as having to have equipment shipped to an employee's home instead of going to the office to pick it up, was unique to the WFH situation. Second, our results show software development teams hold social events to build a social connection with new hires. However, our results do suggest that some new hires do not feel socially connected to their team.  We found many participants responded that they did not have the same opportunities to meet with their peers they would have had if they had been onboarded in the physical office.  Finally, we found new hires desired more meetings with their managers and more overall interaction with their team. %

%% file: discussion.tex
\section{Discussion of Results}\label{sec:results}
Our paper advances the software engineering literature on onboarding in two key directions.  First, we contribute to the literature with empirical evidence of unique challenges that new hires encountered during onboarding in the \covid pandemic.  We surveyed 267 professional developers at Microsoft that had recently onboarded during the pandemic.  We release our survey questions via an online appendix~\cite{rodeghero2020survey}.  We have analyzed the responses to this survey and found that new hires struggled with communication and bonding with their team.  We found that 1:1 meetings can help new hires and that finding useful documentation can be difficult.  These results confirm some of the challenges of a previous study on onboarding~\cite{hemphill2011not}.  However, this work sheds light on the challenges that come from onboarding during a pandemic. Additionally, it is possible that typical remote work challenges, such as documentation, are more impactful and amplified due to other challenges, such as difficulties with communication. The many challenges suggested by our data, such as communication and collaboration challenges, are most likely challenges related to \covid pandemic and WFH orders. This paper focuses on the remote onboarding of developers to teams that worked in a physical location and quickly transitioned to working from home.  This, in itself, is a unique situation for both onboarding and the field of software development.

In addition to our survey, we conducted 8 interviews with new hires and early-career employees at Microsoft.   The confirmatory interviews were conducted to understand the remote onboarding experience further and validate our onboarding survey. We conducted open-ended interviews where we listened to stories of remote onboarding and the challenges that were faced. %

Although this work contributes important evidence of the challenges and needs of new hires onboarding, there are some limitations that we plan to address in future work.  We focused our efforts on exploring the process and experiences that new hires had at Microsoft.  For future work, we plan to explore the remotely onboarded new hires' productivity and job satisfaction. We will investigate how the team connection new remote hires formed with their team impacts their productivity and job satisfaction.  It is important to understand the challenges that new hires, teams that are hiring remote developers, and managers face.  It is also important to note that the type of work currently occurring during the pandemic (e.g., the WFH transition) is not an example of the traditional remote work~\cite{hanselman_2020}.  The difficulties of day to day life are unique (e.g., the lack of human connection due to social distancing) and may greatly impact remote workers and new hires onboarding.  

\section{Recommendations for Onboarding}\label{recs}
\begin{tcolorbox}[width=0.48\textwidth]
\underline{\textbf{Recommendations for Remote Onboarding:}}
\renewcommand{\labelitemi}{$\blacksquare$}
 \begin{itemize}
   \item  Promote communication \& asking for help.
   \item Encourage teams to turn cameras on.
   \item Schedule 1:1 meetings.
   \item Provide information about the organization.
   \item Emphasize team building.
   \item Assign an onboarding buddy.
   \item Assign an onboarding technical mentor.
   \item Support multiple onboarding speeds.
   \item Assign a simple first task.
   \item Provide up-to-date documentation.
   
 \end{itemize} 
\end{tcolorbox}
After collecting the survey data and conducting interviews, the authors discussed our observations and results.  We developed recommendations for both new hires onboarding, managers, and software development teams from this discussion.  We recommend the following tasks to improve the remote onboarding process.

\myparagraph{Promote communication and asking for help.} Some new hires reported that they feel that they are bothering or interrupting their teammates when they reach out to ask a question.  Due to the \covid pandemic, it may be that people are worried about bothering others at home while they juggle work and family.  The anxiety of asking for help may have increased with the work from home transition.  Interviewees reported anxiety around communication.  They reported not knowing when to communicate and how to communicate with their team.  First, we recommend that new hires be provided with guidance and ``best practices'' to reach out to their managers and team members. We also recommend that managers and team members encourage new hires to reach out for help and ask questions when they arise.  Another solution is for managers to meet with new hires daily for short periods of time to make sure that they can ask questions.   %

\myparagraph{Encourage team members to turn cameras on.} We found that multiple participants desired their teammates to turn on their video cameras during video calls.  They reported that this would help them better understand the dynamics of the team.  They also mentioned that it would help them bond and form connections with their peers.  We recommend that managers encourage teams to turn their cameras on in all meetings.  Additionally, we highly encourage and recommend that managers also turn on their cameras to set an example for their team.

\myparagraph{Schedule 1:1 meetings.} Both survey participants and interviewees reported that 1:1 meetings with both their teammates and manager were helpful.  We also found that new hires desired short but frequent, one on one meetings throughout the week.  We recommend that managers have formalized and consistent short daily meetings with new hires to ensure that they are not blocked on tasks.  In the interviews, we found that managers found their new hires' onboarding process to be more positive if they had formalized, consistent, and sustained 1:1 interactions.  In addition to the manager holding 1:1s, we also recommend that team members reach out to new hires and meet with them individually.  This may help new hires feel more comfortable asking for  help and feel more part of the team. We also found that participants desired to meet with their \emph{skip-level manager} (the new hire's manager's manager).  We recommend that managers assist in setting up a meeting between new hires and the skip-level manager, which helps to better understand the company's strategy, organization, and culture.

\myparagraph{Provide information about the organization.} We found that the new hires reported that they did not fully understand the company's organization.  By understanding the broader business context and opportunities of the organization, so they could understand how their individual role and their team's role fit within that.  We recommend that managers provide information to new hires early on about the company and team organization.  

\myparagraph{Emphasize team building.} Based upon both the survey responses and interviews, we found new hires desired a stronger social connection with their team.  We recommend having regular social activities for team bonding and, later, for maintaining social team connection. Team connection is directly related to job satisfaction and productivity~\cite{boland2004transitioning}.  Therefore, we emphasize the need for new hires to build strong connections to their team.  These activities should be inclusive.  This includes teams with hybrid workers where some team members are remote while others are in person.  Due to the current pandemic, we recommend holding activities virtually since most developers are working remotely.  This will allow all team members to participate. Some examples of remote team activities we recommend include happy hours, a workout class such as yoga, coffee chats, or playing online games together.  

\myparagraph{Assign an onboarding buddy.} New hires found that their onboarding buddy was helpful to their onboarding.\myquote{My manager and new hire buddy was very helpful in getting me oriented with the team and the work.}{R_3PcXJn7Sd0VIohw} However, those who did not have an onboarding buddy suggested that having one could improve the onboarding process.
The onboarding buddy's goal is to help new hires find resources they need, connect the new hire with others within the company, and be a source for asking questions.  The onboarding buddy helps to see that the new hire's process is smooth and ensures that the new hire feels welcome.  We recommend that the onboarding buddy does not come from within the new hire's direct team but rather a team in the same organization.  When the buddy is outside of the direct team, the buddy can help the new hire network outside the new hire's team, discuss career growth, and any problems during the onboarding process. 

\myparagraph{Assign an onboarding technical mentor.} We recommend that each new hire is assigned an onboarding technical mentor.  The technical mentor is a mentor that helps new hires with the low-level details of their work.  The mentor should come from within the new hire's team and assist the new hire with technical issues (e.g., code questions). We recommend that the mentor be on the same team and work with the code for a significant period of time so that they can answer specific detailed questions.  This mentor should be available to assist and respond promptly.  We note that mentors need to sustain long enough connections with the new hire and avoid dropping out of the mentor role too early.  By making the mentor role a normal job function, managers can alleviate the anxiety around sending multiple and frequent questions.  

\myparagraph{Support multiple onboarding speeds.} From the interviews, we found that some hires may prefer different onboarding or ramping-up speeds.  Because each new hire has a unique background before joining a new team or company, their needs may differ from their onboarding peers concurrently.  For example, a new hire from college may have less workflow experience than an industry hire.  Industry hires may quickly join a new team and create a pull request in the first week.  They may need less guidance.  However, this is not necessarily the case and may need to be evaluated on a case by case basis.  
Organizations could offer a normal onboarding track and a fast track, and new hires could switch between tracks as they get more familiar with the company.
We believe new hires will have a higher level of job satisfaction if the onboarding process is not a one-size-fits-all process.

\myparagraph{Assign a simple first task.} The first task that new hires work on can be daunting.  New hires need to learn how the software functions, the architecture of the project they have been assigned to, the coding and documentation standards that the team follows, the workflow, and much more.  We found that new hires often take a few weeks to create their first pull request.  This means that they may not understand the workflow quickly and, therefore, they may not understand some of the discussions had during team meetings until after they have gone through the process of creating their first pull request.  We recommend new hires complete a simple task, such as fixing a spelling error. This allows new hires to quickly go through the process of building a project, creating a pull request, submitting it, and having it reviewed.

\myparagraph{Provide up-to-date documentation.} We found that new hires that had ample documentation reported that they had a smooth onboarding process.  These new hires were able to find the resources that they needed. However, we found that many new hires found the documentation to be overwhelming.  Others reported that documentation was many times outdated or nonexistent.  The survey responses on documentation were similar to the findings in our interviews.  It is important to note that new hires need context to understand the documentation.  They need up-to-date documentation, and they need to know where they fit into the project in relation to the documentation. With remote work, because communication may be delayed and questions may not be answered as quickly to the traditional office work, ideally, documentation provides a resource for remote hires to access easily.  For example, one survey participant talked about the challenges they encountered with the documentation.

  \myquote{There are a lot of small hitches in the documentation for setting up the dev environment, a lot of small ways things can go wrong, that are the sort that can be easily fixed by finding anyone who has ever done them before and going [to ask questions] at them. When those people are all busy and miles away, and you need to interrupt them by texting them while their Teams status is red, that gets harder.}{R_3nMiXSu3jGTAXlD}
  Finally, we also recommend providing documentation that provides new hires with workflow information, allowing new hires to understand the team's day-to-day language.

\medskip
To summarize our recommendations, we recommend creating a clear onboarding process that is easily modifiable based on the new hires' needs.  The process should integrate important technical aspects early on (e.g., introducing the workflow) and social aspects (e.g., team bonding activities).  We also suggest providing new hires with both an onboarding mentor and a technical mentor.  In general, due to the pandemic and the shift to remote work, we suggest building a culture of allyship for remote workers~\cite{rodeghero2020empowering}.  The recommendations that we provide are primarily for onboarding new hires remotely due to the current crisis.  We note that this work has focused primarily on new remote hires.  We know that the future of work has drastically changed~\cite{press_2020}. We believe that these recommendations will help those who continue to hire during the current pandemic and remote developers in general. Finally, we believe that these recommendations (e.g., team bonding activities and co-working) will also benefit all remote workers and teams as a whole. 

%% file: conclusion.tex
\section{Conclusion}
We have presented a study of remote onboarding during the \covid pandemic.  We surveyed 267 professional developers at Microsoft. We explored several research questions to understand the remote onboarding process and the challenges that new hires faced during the process. We showed the challenges newly hired developers encountered included difficulties finding documentation, communication, asking for help, and bonding with teammates.  We also explored the new hires' connections within their teams and how teams bonded. Finally, we provide recommendations to assist managers and teams in developing a remote onboarding process during a crisis.  

In future work, we plan to explore the importance of each challenge presented in our results.  We believe that some of the challenges may be more important to new hires and teams than others.  Additionally, we will explore onboarding remotely from the managers' side of view and their challenges.  We also plan to implement interventions based upon our findings to explore methodologies that may decrease the challenges that both new hires and teams have faced during WFH.

%% file: conference_101719.bbl
\begin{thebibliography}{10}
\providecommand{\url}[1]{#1}
\csname url@samestyle\endcsname
\providecommand{\newblock}{\relax}
\providecommand{\bibinfo}[2]{#2}
\providecommand{\BIBentrySTDinterwordspacing}{\spaceskip=0pt\relax}
\providecommand{\BIBentryALTinterwordstretchfactor}{4}
\providecommand{\BIBentryALTinterwordspacing}{\spaceskip=\fontdimen2\font plus
\BIBentryALTinterwordstretchfactor\fontdimen3\font minus
  \fontdimen4\font\relax}
\providecommand{\BIBforeignlanguage}[2]{{%
\expandafter\ifx\csname l@#1\endcsname\relax
\typeout{** WARNING: IEEEtran.bst: No hyphenation pattern has been}%
\typeout{** loaded for the language `#1'. Using the pattern for}%
\typeout{** the default language instead.}%
\else
\language=\csname l@#1\endcsname
\fi
#2}}
\providecommand{\BIBdecl}{\relax}
\BIBdecl

\bibitem{labuschagne2015onboarding}
A.~Labuschagne and R.~Holmes, ``Do onboarding programs work?'' in \emph{Mining
  Software Repositories (MSR), 2015 IEEE/ACM 12th Working Conference on}.\hskip
  1em plus 0.5em minus 0.4em\relax IEEE, 2015, pp. 381--385.

\bibitem{bonar1983uncovering}
J.~Bonar and E.~Soloway, ``Uncovering principles of novice programming,'' in
  \emph{Proceedings of the 10th ACM SIGACT-SIGPLAN symposium on Principles of
  programming languages}.\hskip 1em plus 0.5em minus 0.4em\relax ACM, 1983, pp.
  10--13.

\bibitem{williams2004initial}
L.~Williams, A.~Shukla, and A.~I. Anton, ``An initial exploration of the
  relationship between pair programming and brooks' law,'' in \emph{Agile
  Development Conference, 2004}.\hskip 1em plus 0.5em minus 0.4em\relax IEEE,
  2004, pp. 11--20.

\bibitem{fagerholm2014role}
F.~Fagerholm, A.~S. Guinea, J.~M{\"u}nch, and J.~Borenstein, ``The role of
  mentoring and project characteristics for onboarding in open source software
  projects,'' in \emph{Proceedings of the 8th ACM/IEEE international symposium
  on empirical software engineering and measurement}.\hskip 1em plus 0.5em
  minus 0.4em\relax ACM, 2014, p.~55.

\bibitem{steinmacher2015social}
I.~Steinmacher, T.~Conte, M.~A. Gerosa, and D.~Redmiles, ``Social barriers
  faced by newcomers placing their first contribution in open source software
  projects,'' in \emph{Proceedings of the 18th ACM conference on Computer
  supported cooperative work \& social computing}, 2015, pp. 1379--1392.

\bibitem{johnson2010learning}
M.~Johnson and M.~Senges, ``Learning to be a programmer in a complex
  organization: A case study on practice-based learning during the onboarding
  process at google,'' \emph{Journal of Workplace Learning}, vol.~22, no.~3,
  pp. 180--194, 2010.

\bibitem{sim1998ramp}
S.~E. Sim and R.~C. Holt, ``The ramp-up problem in software projects: A case
  study of how software immigrants naturalize,'' in \emph{Software Engineering,
  1998. Proceedings of the 1998 International Conference on}.\hskip 1em plus
  0.5em minus 0.4em\relax IEEE, 1998, pp. 361--370.

\bibitem{zhou2010developer}
M.~Zhou and A.~Mockus, ``Developer fluency: Achieving true mastery in software
  projects,'' in \emph{Proceedings of the eighteenth ACM SIGSOFT international
  symposium on Foundations of software engineering}.\hskip 1em plus 0.5em minus
  0.4em\relax ACM, 2010, pp. 137--146.

\bibitem{peterson_2017}
\BIBentryALTinterwordspacing
B.~Peterson, ``Travis kalanick lasted in his role for 6.5 years - five times
  longer than the average uber employee,'' Aug 2017. [Online]. Available:
  \url{http://www.businessinsider.com/employee-retention-rate-top-tech-companies-2017-8}
\BIBentrySTDinterwordspacing

\bibitem{hill1998influences}
E.~J. Hill, B.~C. Miller, S.~P. Weiner, and J.~Colihan, ``Influences of the
  virtual office on aspects of work and work/life balance,'' \emph{Personnel
  psychology}, vol.~51, no.~3, pp. 667--683, 1998.

\bibitem{chowell2020covid}
G.~Chowell and K.~Mizumoto, ``The covid-19 pandemic in the usa: what might we
  expect?'' \emph{The Lancet}, vol. 395, no. 10230, pp. 1093--1094, 2020.

\bibitem{brownlee_2020}
D.~Brownlee, ``Twitter, square announce work from home forever option: What are
  the risks?''
  https://www.forbes.com/sites/danabrownlee/2020/05/18/twitter-square-announce-work-from-home-forever-optionwhat-are-the-risks/,
  May 2020.

\bibitem{britto2018onboarding}
R.~Britto, D.~S. Cruzes, D.~Smite, and A.~Sablis, ``Onboarding software
  developers and teams in three globally distributed legacy projects: A
  multi-case study,'' \emph{Journal of Software: Evolution and Process},
  vol.~30, no.~4, p. e1921, 2018.

\bibitem{gruman2006organizational}
J.~A. Gruman, A.~M. Saks, and D.~I. Zweig, ``Organizational socialization
  tactics and newcomer proactive behaviors: An integrative study,''
  \emph{Journal of vocational behavior}, vol.~69, no.~1, pp. 90--104, 2006.

\bibitem{klein2015specific}
H.~J. Klein, B.~Polin, and K.~Leigh~Sutton, ``Specific onboarding practices for
  the socialization of new employees,'' \emph{International Journal of
  Selection and Assessment}, vol.~23, no.~3, pp. 263--283, 2015.

\bibitem{lynch2010reducing}
K.~Lynch and G.~Buckner-Hayden, ``Reducing the new employee learning curve to
  improve productivity,'' \emph{Journal of healthcare risk management},
  vol.~29, no.~3, pp. 22--28, 2010.

\bibitem{johnson2005get}
L.~Johnson, ``Get your new managers moving,'' \emph{Harvard Management Update},
  vol.~10, no.~6, 2005.

\bibitem{moyson2018organizational}
S.~Moyson, N.~Raaphorst, S.~Groeneveld, and S.~Van~de Walle, ``Organizational
  socialization in public administration research: A systematic review and
  directions for future research,'' \emph{The American Review of Public
  Administration}, vol.~48, no.~6, pp. 610--627, 2018.

\bibitem{begel2008pair}
A.~Begel and N.~Nagappan, ``Pair programming: what's in it for me?'' in
  \emph{Proceedings of the Second ACM-IEEE international symposium on Empirical
  software engineering and measurement}.\hskip 1em plus 0.5em minus 0.4em\relax
  ACM, 2008, pp. 120--128.

\bibitem{hemphill65not}
L.~Hemphill and A.~Begel, ``Not seen and not heard: Onboarding challenges in
  newly virtual teams,'' \emph{131.107}, vol.~65, 2011.

\bibitem{ford2019remotework}
\BIBentryALTinterwordspacing
D.~Ford, R.~Milewicz, and A.~Serebrenik, ``How remote work can foster a more
  inclusive environment for transgender developers,'' in \emph{Proceedings of
  the 2nd International Workshop on Gender Equality in Software Engineering},
  ser. GE '19.\hskip 1em plus 0.5em minus 0.4em\relax IEEE Press, 2019, p.
  9–12. [Online]. Available: \url{https://doi.org/10.1109/GE.2019.00011}
\BIBentrySTDinterwordspacing

\bibitem{mazzina2017remotefirst}
A.~Mazzina, ``What it means to be a remote-first company,'' February 2017,
  retrieved June 4, 2020 from
  \url{https://stackoverflow.blog/2017/02/08/means-remote-first-company/}.

\bibitem{gitlab2020remotereport}
GitLab, ``The remote work report by gitlab: The future of work is remote,''
  March 2020, retrieved June 4, 2020 from
  \url{https://page.gitlab.com/rs/194-VVC-221/images/the-remote-work-report-by-gitlab.pdf}.

\bibitem{gitlab2020playbook}
------, ``The remote playbook from the largest all-remote company in the
  world,'' March 2020, retrieved June 4, 2020 from
  \url{https://about.gitlab.com/resources/downloads/ebook-remote-playbook.pdf}.

\bibitem{pardue2017remotework}
J.~Pardue, ``Making remote work: Behind the scenes at stack overflow,''
  September 2017, retrieved June 4, 2020 from
  \url{https://stackoverflow.blog/2017/09/29/making-remote-work-behind-scenes/}.

\bibitem{hanselman_2020}
\BIBentryALTinterwordspacing
S.~Hanselman, ``Quarantine work !== remote work. i've been working remotely
  with success for 13 years, and i've never been close to burn out. i've been
  working quarantined for over a month and i'm feeling a tinge if burn out for
  the first time in my life. take care of yourself folks. really.'' Apr 2020.
  [Online]. Available:
  \url{https://twitter.com/shanselman/status/1252040170783641600}
\BIBentrySTDinterwordspacing

\bibitem{usher2020life}
K.~Usher, N.~Bhullar, and D.~Jackson, ``Life in the pandemic: Social isolation
  and mental health,'' \emph{Journal of Clinical Nursing}, 2020.

\bibitem{cluver2020parenting}
L.~Cluver, J.~M. Lachman, L.~Sherr, I.~Wessels, E.~Krug, S.~Rakotomalala,
  S.~Blight, S.~Hillis, G.~Bachmand, O.~Green \emph{et~al.}, ``Parenting in a
  time of covid-19,'' \emph{The Lancet}, 2020.

\bibitem{horesh2020traumatic}
D.~Horesh and A.~D. Brown, ``Traumatic stress in the age of covid-19: A call to
  close critical gaps and adapt to new realities.'' \emph{Psychological Trauma:
  Theory, Research, Practice, and Policy}, vol.~12, no.~4, p. 331, 2020.

\bibitem{bao2020does}
L.~Bao, T.~Li, X.~Xia, K.~Zhu, H.~Li, and X.~Yang, ``How does working from home
  affect developer productivity? -- a case study of baidu during covid-19
  pandemic,'' 2020.

\bibitem{ralph2020pandemic}
P.~Ralph, S.~Baltes, G.~Adisaputri, R.~Torkar, V.~Kovalenko, M.~Kalinowski,
  N.~Novielli, S.~Yoo, X.~Devroey, X.~Tan, M.~Zhou, B.~Turhan, R.~Hoda,
  H.~Hata, G.~Robles, A.~M. Fard, and R.~Alkadhi, ``Pandemic programming: How
  covid-19 affects software developers and how their organizations can help,''
  2020.

\bibitem{ford2020tale}
D.~Ford, M.-A. Storey, T.~Zimmermann, C.~Bird, S.~Jaffe, C.~Maddila, J.~L.
  Butler, B.~Houck, and N.~Nagappan, ``A tale of two cities: Software
  developers working from home during the covid-19 pandemic,'' 2020.

\bibitem{reagans2001networks}
R.~Reagans and E.~W. Zuckerman, ``Networks, diversity, and productivity: The
  social capital of corporate r\&d teams,'' \emph{Organization science},
  vol.~12, no.~4, pp. 502--517, 2001.

\bibitem{yilmaz2016effective}
M.~Yilmaz, R.~V. O'Connor, and P.~Clarke, ``Effective social productivity
  measurements during software development—an empirical study,''
  \emph{International Journal of Software Engineering and Knowledge
  Engineering}, vol.~26, no.~03, pp. 457--490, 2016.

\bibitem{ayazlar2014effect}
G.~Ayazlar and B.~G{\"u}zel, ``The effect of loneliness in the workplace on
  organizational commitment,'' \emph{Procedia-Social and Behavioral Sciences},
  vol. 131, pp. 319--325, 2014.

\bibitem{lopes2005informal}
R.~Lopes~Morrison, ``Informal relationships in the workplace: Associations with
  job satisfaction, organisational commitment and turnover intentions,'' Ph.D.
  dissertation, Massey University, 2005.

\bibitem{dery2016seeing}
K.~Dery and E.~Hafermalz, ``Seeing is belonging: Remote working, identity and
  staying connected,'' in \emph{The impact of ICT on work}.\hskip 1em plus
  0.5em minus 0.4em\relax Springer, 2016, pp. 109--126.

\bibitem{mayring2004qualitative}
P.~Mayring, ``Qualitative content analysis,'' \emph{A companion to qualitative
  research}, vol.~1, no. 2004, pp. 159--176, 2004.

\bibitem{spencer2004card}
D.~Spencer and T.~Warfel, ``Card sorting: a definitive guide,'' \emph{Boxes and
  arrows}, vol.~2, pp. 1--23, 2004.

\bibitem{garrison2006revisiting}
D.~R. Garrison, M.~Cleveland-Innes, M.~Koole, and J.~Kappelman, ``Revisiting
  methodological issues in transcript analysis: Negotiated coding and
  reliability,'' \emph{The Internet and Higher Education}, vol.~9, no.~1, pp.
  1--8, 2006.

\bibitem{garousi2013evaluating}
G.~Garousi, V.~Garousi, M.~Moussavi, G.~Ruhe, and B.~Smith, ``Evaluating usage
  and quality of technical software documentation: an empirical study,'' in
  \emph{Proceedings of the 17th International Conference on Evaluation and
  Assessment in Software Engineering}, 2013, pp. 24--35.

\bibitem{lethbridge2003software}
T.~C. Lethbridge, J.~Singer, and A.~Forward, ``How software engineers use
  documentation: The state of the practice,'' \emph{IEEE software}, vol.~20,
  no.~6, pp. 35--39, 2003.

\bibitem{rodeghero2020survey}
\BIBentryALTinterwordspacing
P.~Rodeghero, D.~Ford, and T.~Zimmermann, ``Survey on onboarding during a
  pandemic,'' Microsoft, Tech. Rep. MSR-TR-2020-36, October 2020. [Online].
  Available:
  \url{https://www.microsoft.com/en-us/research/publication/survey-on-onboarding-during-a-pandemic/}
\BIBentrySTDinterwordspacing

\bibitem{hemphill2011not}
L.~Hemphill and A.~Begel, ``Not seen and not heard: Onboarding challenges in
  newly virtual teams,'' \emph{131.107}, vol.~65, 2011.

\bibitem{boland2004transitioning}
D.~Boland and B.~Fitzgerald, ``Transitioning from a co-located to a
  globally-distributed software development team: A case study at analog
  devices inc,'' in \emph{3rd Workshop on Global Software Development}.\hskip
  1em plus 0.5em minus 0.4em\relax IET, 2004.

\bibitem{rodeghero2020empowering}
P.~Rodeghero and T.~Hernandez, ``Empowering and supporting remote software
  development team members through a culture of allyship,'' August 2020,
  microsoft Research Symposium on the New Future of Work.

\bibitem{press_2020}
\BIBentryALTinterwordspacing
G.~Press, ``The future of work post-covid-19,'' Jul 2020. [Online]. Available:
  \url{https://www.forbes.com/sites/gilpress/2020/07/15/the-future-of-work-post-covid-19/}
\BIBentrySTDinterwordspacing

\end{thebibliography}
